\documentclass[reprint, superscriptaddress, amsmath, amssymb, aps, pra, nolongbibliography]{revtex4-2}
\usepackage{graphicx,xcolor}
\usepackage{dsfont,braket,siunitx,upgreek}
\definecolor{darkblue}{rgb}{0,0.3,0.7}
\usepackage{hyperref, comment}
\hypersetup{
colorlinks=true,
linkcolor=darkblue,
filecolor=blue,
citecolor=darkblue,  
urlcolor=darkblue,
}

\begin{document}

\preprint{APS/123-QED}

\title{From broadband biphotons to frequency combs via\\spectral compression with time-varying cavities}
\author{Karthik V. Myilswamy}
\thanks{equal contribution.}
\author{Jordan A. Gaines}
\thanks{equal contribution.}
\author{Jason D. McKinney}
\affiliation{School of Electrical and Computer Engineering and Purdue Quantum Science and Engineering Institute, Purdue University, West Lafayette, Indiana 47907, USA}
\author{Joseph M. Lukens}
\email{joseph.lukens@asu.edu}
\affiliation{Quantum Information Science Section, Oak Ridge National Laboratory, Oak Ridge, Tennessee 37831, USA}
\affiliation{Research Technology Office and Quantum Collaborative, Arizona State University, Tempe, Arizona 85287, USA}
\author{Andrew M. Weiner}
\affiliation{School of Electrical and Computer Engineering and Purdue Quantum Science and Engineering Institute, Purdue University, West Lafayette, Indiana 47907, USA}

\date{\today}

\begin{abstract}
Biphoton frequency combs are promising resources for quantum networking due in large part to their compatibility with the telecommunication infrastructure. In this work, we propose a method to periodically compress broadband frequency-entangled photons into biphoton frequency combs by utilizing time-varying linear cavities. 
Our approach hinges on rapid modulation of the input cavity coupling, yielding high spectral purity in each output comb line similar to that achieved with narrowband filters, but without the associated loss in flux. 
We examine the dependence of spectral purity and compression on coupling strength, cavity loss, and switching speed, finding realistic regimes supporting purities in excess of 0.999 and peak enhancement factors of 100$\times$ and beyond.
\end{abstract}

\maketitle

Recently, encoding quantum information as a coherent superposition of multiple discrete frequencies---in states known as quantum frequency combs---has attracted significant attention due to advantages including fiber-optic compatibility and the support of high-dimensional quantum information processing~\cite{kues2019quantum,lu2022bayesian,lu2023frequency}. Broadband frequency-entangled photons can be generated through spontaneous parametric downconversion (SPDC) in $\chi^{\left(2\right)}$ bulk crystals and waveguides and then converted into biphoton frequency combs (BFCs) using 
Fabry--Perot (FP) cavities. Additionally, these broadband biphotons (either as a comb or not) can function as heralded single-photon sources by detecting one of the paired photons~\cite{ramelow2013highly}. However, the time-energy correlations inherent in broadband SPDC reduce spectral purity, or equivalently, spectral factorability~\cite{grice2001eliminating, mosley2008heralded,laiho2009producing,laudenbach2016modelling}, diminishing photon indistinguishability and impairing the interference between independent photons critical for quantum networking operations like entanglement swapping~\cite{merkouche2022heralding}.

Narrowband filters are often employed to eliminate frequency correlations in broadband SPDC, but this approach significantly reduces the output flux, as nonresonant portions of the spectrum are discarded~\cite{laiho2009producing,meyer2017limits}.
Following seminal work to generate spectrally pure photons directly via
pulsed pumping in bulk crystals that achieve group velocity matching
~\cite{grice2001eliminating, mosley2008heralded}, recent years have witnessed a growth of strategies incorporating device engineering to generate factorable biphotons with high flux and purity, 
focusing on the nonlinearity profile~\cite{branczyk2010engineered}, poling patterns and periods~\cite{chen2017efficient,graffitti2017pure, graffitti2018independent, pickston2021optimised}, and waveguide dispersion~\cite{xin2022spectrally}.

With the rise of integrated photonics, resonant structures like microrings have emerged as popular BFC sources, which avoid the flux loss associated with post-generation filtering. These sources typically leverage spontaneous four-wave mixing (SFWM), where the achievable spectral purity is limited to a theoretical maximum of 0.93~\cite{vernon2017truly, vaidya2020broadband, myilswamy2023time}, unless additional techniques are used, such as engineering the input coupling~\cite{liu2019high, burridge2023integrate, borghi2024uncorrelated, alexander2024manufacturable} or shaping pump pulses~\cite{christensen2018engineering, burridge2020high}.
Accordingly, existing strategies for realizing spectrally separable biphotons can be broadly separated into two classes: (i) frequency-entangled biphoton generation followed by fixed spectral filters and (ii) direct generation via engineering of the pump and nonlinear medium. The first family achieves high purity straightforwardly but at the cost of photon flux, while the second group is generally more complex but is in principle lossless. 

In this work, we present a fundamentally different approach that bridges both approaches---the post-generation nature of a filter with the flux-saving abilities of state engineering. We propose periodic spectral compression of broadband biphotons into BFCs using linear optical cavities with time-variant input coupling. 
Previously, we detailed the prospects for such time-varying cavities to compress the spectrum and shape the temporal mode of broadband single photons for enhanced interactions with quantum memories~\cite{myilswamy2020spectral, myilswamy2022temporal}.
Here, we extend this concept to a broader framework, utilizing impulse response formulations to apply it to two-photon systems. 
Through detailed simulations with a variety of parameters, we show how this approach can convert broadband biphotons with a continuous spectrum into BFCs with highly pure frequency bins ($>$0.999), along with peak enhancement factors $>$10$^4$ ($>$10$^2$) for 0~dB (0.2~dB) roundtrip loss and minimal degradation for finite switching times on the order of the cavity roundtrip.

\begin{figure}[tb!]
\centering\includegraphics[trim=0 120 65 0,clip,width=3.5in]{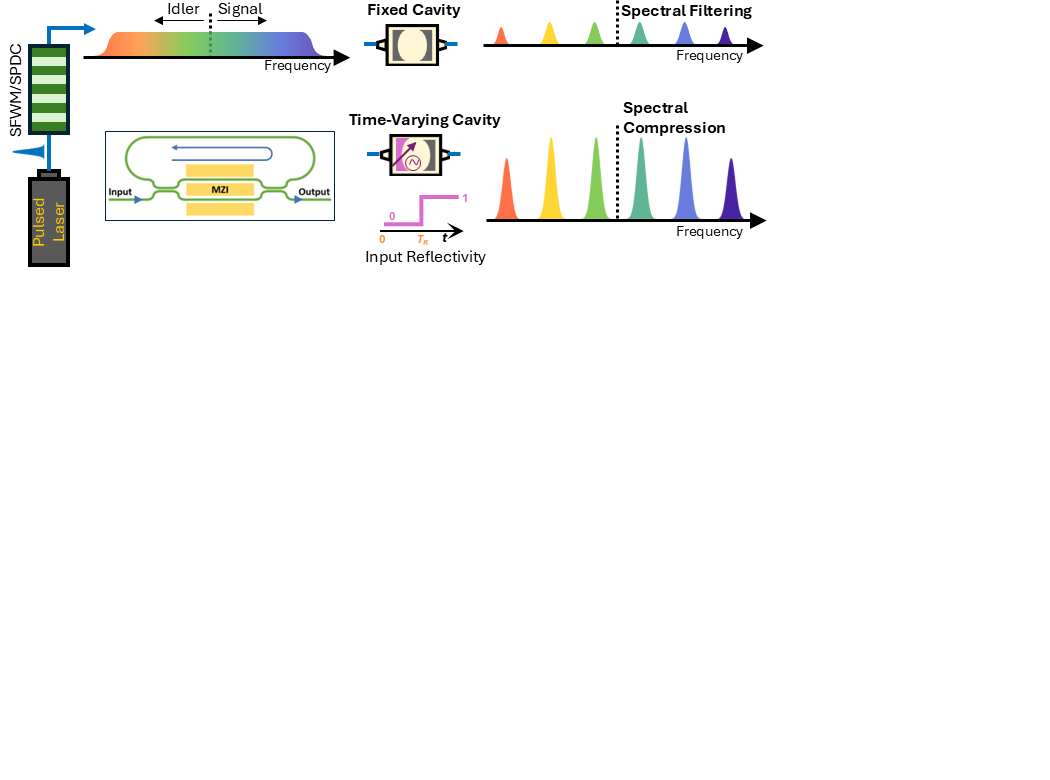}
\caption{Biphoton compression scheme. Whereas a broadband biphoton generated from pulsed SPDC or SFWM can be converted into a BFC with loss through a fixed cavity filter (top), a time-varying cavity can achieve periodic compression without losing energy (bottom). 
The box shows a possible integrated implementation.}
\label{fig1}
\end{figure}

Figure~\ref{fig1} illustrates our approach. 
A broadband biphoton generated from pulsed SPDC or SFWM is fully captured within the time-varying cavity by rapidly switching the input reflectivity from $0$ to $1$ as the biphoton enters. Once inside, the biphoton eventually exits through the partially reflecting output mirror, producing a temporal wavefunction with multiple replicas decreasing exponentially in amplitude. These replicas, with fixed relative delays corresponding to multiples of the cavity roundtrip time $T_R$ for both the signal and idler, achieve spectral compression through interference. Unlike fixed cavities that discard nonresonant portions, the biphoton energy is compressed into narrowband resonances, improving spectral purity while eliminating the typical loss (represented pictorially in Fig.~\ref{fig1} by the difference in peak heights of the two output spectra).
The box in Fig.~\ref{fig1} shows an integrated photonic time-varying cavity, featuring a microring resonator with a Mach--Zehnder interferometer (MZI) for tunable coupling~\cite{xue2022breaking, jia2023electrically, herrmann2024arbitrary}. Thin-film lithium niobate (TFLN) is an optimal platform for this design, offering both low-loss waveguides and fast-switching MZIs~\cite{boes2023lithium}. 
For the majority of the discussion, we focus on the FP cavity as a representative example, but all findings readily extend to the microring version as well~\cite{myilswamy2020spectral,myilswamy2022temporal}.

A linear time-\emph{variant} system is characterized by a two-time impulse response function $h(t,\tau)$, which describes the system's output in response to an impulse $\delta(t-\tau)$ at time $\tau$. From a frequency domain viewpoint, the Fourier transform 
$H(\omega,\omega^{\prime}) = \frac{1}{2\pi} \iint h(t,\tau) e^{-i(\omega t - \omega' \tau)} d\tau dt$
describes the transfer of input frequency content at $\omega^{\prime}$ to the output frequency $\omega$ ~\cite{zadeh1950frequency, xiao2011spectral}.
This frequency transfer, facilitated by the time-varying nature of the system, forms the basis of spectral compression. 
Thus, the input-output relations in the time and frequency domains are expressed as $y(t) = \int h(t,\tau) x(\tau) d\tau$ and $Y(\omega) = \int H(\omega,\omega^{\prime}) X(\omega^{\prime}) d\omega^{\prime}$, where the input (output) signals satisfy $X(\omega)= \int x(t) e^{-i\omega t} dt$ $\left(Y(\omega)= \int y(t) e^{-i\omega t} dt\right)$.

For a time-varying cavity where the input (field) reflection coefficient $r_1$ rapidly switches from 0 to 1 at $t=T_R$ while the output $r_2$ remains fixed, the impulse response becomes
\begin{equation}
\small
    h(t,\tau) = \left\{ \begin{array}{lcr}
  \sum_{m=0}^{\infty} r_2^mt_2  \delta \left(t-\tau-\left[m+\frac{1}{2}\right]T_R\right)  &:& 0\leq\tau \leq T_R \\
  0 &:&\ \text{otherwise},
\end{array} \right.
\label{eq-h_t_tau}
\end{equation}
\begin{equation}
\small
    H(\omega,\omega^\prime) = \frac{T_R}{2\pi} \frac{t_2e^{-i\omega\frac{T_R}{2}}}{1-r_2e^{-i\omega T_R}} e^{-i(\omega-\omega^\prime)\frac{T_R}{2}}\mathrm{sinc}\left[(\omega-\omega^\prime)\frac{T_R}{2}\right],
    \label{eq-H_W_W'}
\end{equation}
where we assume for convenience $r_1 = 1$ for $t<0$, which has no impact on the input pulses, as they are confined between $t=0$ and $t=T_R$. Intuitively, Eq.~(\ref{eq-h_t_tau}) relates how the output at any given time $t$ consists of a superposition of inputs that have bounced off the output mirror $m$ times before exiting through the field transmission $t_2=\sqrt{1-|r_2|^2}$, while Eq.~(\ref{eq-H_W_W'}) highlights how this interference results in a nontrivial frequency translation.

\begin{figure}[tb!]
\centering\includegraphics[width=3.5in]{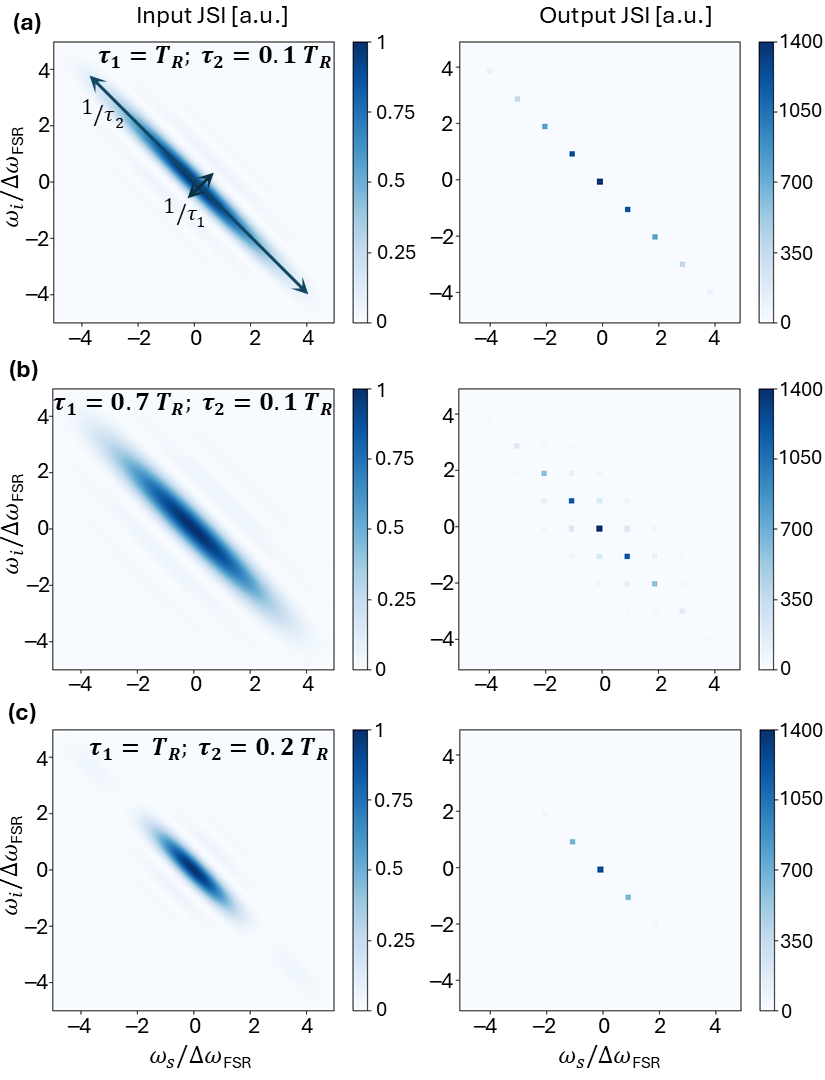}
\caption{Spectral compression of broadband biphotons into BFCs. The input (output) JSIs appear in the left (right) column, for the following cases: (a) $(\tau_1,\tau_2)=(T_R,0.1T_R)$, (b) $(\tau_1,\tau_2)=(0.7T_R,0.1T_R)$, and 
(c) $(\tau_1,\tau_2)=(T_R,0.2T_R)$.}
\label{fig2}
\end{figure}

The input biphoton is characterized by a complex spectral wavefunction $\psi_\mathrm{in}(\omega_s,\omega_i)$, where $\omega_s$ ($\omega_i$) represents the baseband signal (idler) frequency. Since both photons traverse the same cavity in our method, the output biphoton wavefunction $\psi_\mathrm{out}(\omega_s,\omega_i)$ is given by
\begin{equation}
\small
        \psi_\mathrm{out}(\omega_s,\omega_i) = \int_{-\infty}^{\infty}\int_{-\infty}^{\infty} H(\omega_s,\omega_s^\prime) H(\omega_i,\omega_i^\prime) \psi_\mathrm{in}(\omega_s^\prime,\omega_i^\prime) d\omega_s^\prime d\omega_i^\prime.
\end{equation}
For concreteness, we consider input biphotons with $\psi_\mathrm{in}(\omega_s,\omega_i) = \mathrm{sinc}[(\omega_s+\omega_i)\tau_1/2]\mathrm{sinc}[(\omega_s-\omega_i)\tau_2/2] e^{-i(\omega_s+\omega_i)\tau_{s}}$, normalized to a peak of unity for convenience, and with a specific functional form chosen for its time-limitedness---i.e., the time-domain wavepacket is zero for all times outside of $\left[\tau_s-\frac{\tau_1}{2},\tau_s+\frac{\tau_1}{2}\right]$. 
This expression accounts for the main features of a biphoton wavepacket while allowing us to focus on the effect of the cavity itself without complications from decaying temporal tails.
In the frequency domain, $\tau_1$ and $\tau_2$ determine the spread of the biphoton wavefunction along the antidiagonal and diagonal axes, respectively, as shown in Fig.~\ref{fig2}(a). 
We choose $\tau_s = T_R/2$ to center the input between $t=0$ and $t=T_R$; $r_2=0.95$ is assumed. We compute the integrals numerically using Monte Carlo integration with uniform sampling---a simple parallelizable approach for greater computational efficiency than traditional Riemann methods, particularly in high dimensions~\cite{geweke1996monte}. 

First, we examine the case $(\tau_1,\tau_2)=(T_R,0.1T_R)$. 
The input and output joint spectral intensities (JSIs) $|\psi_\mathrm{in}|^2$ and $|\psi_\mathrm{out}|^2$ 
are displayed in the left and right columns of Fig.~\ref{fig2}(a), respectively. For clarity, a coarser plotting grid is used on the output to enhance visibility of the compressed peaks. 
The input biphoton is compressed into multiple resonances along the diagonal, spaced by the free spectral range (FSR) $\Delta\omega_\mathrm{FSR}=2\pi/T_R$. The JSI tapers off away from the spectral center due to the finite bandwidth of the input biphoton. The bin intensities are significantly enhanced compared to the input. The peak value of the bin located at $(\omega_s,\omega_i)/\Delta\omega_\mathrm{FSR} = (0,0)$ is approximately $1400$---i.e., a thousand-fold increase compared to a fixed FP cavity, where the peak value would have been just 1. This increase in JSI peaks, combined with periodic spectral binning, clearly demonstrates the spectral compression effect caused by the time-varying nature of the cavity. 
To assess the extent to which energy is confined to the diagonal as designed, we define the coincidence-to-accidental ratio (CAR) as the ratio of the peak values between the frequency bins at $(\omega_s,\omega_i)/\Delta\omega_\mathrm{FSR} = (0,0)$ and $(0,1)$: the CAR is roughly 350 for the BFC in Fig.~\ref{fig2}(a). (For all CAR values in this paper, we focus on the ratio between these two specific frequency bins in order to fairly assess biphotons with varying bandwidths.)

In the second case, we consider $(\tau_1,\tau_2) = (0.7T_R,0.1T_R)$ 
[Fig.~\ref{fig2}(b)]. The diagonal peaks are roughly similar to those in the first case. However, broader antidiagonal bandwidth leaves significant energy in off-diagonal locations, leading to a CAR of $\sim$8.
Finally, we analyze the case where $(\tau_1,\tau_2) = (T_R,0.2T_R)$ 
[Fig.~\ref{fig2}(c)]. Compared to the first case, the input JSI is narrower along the diagonal but the same along the antidiagonal. This results in fewer strong diagonal output peaks, with a CAR of approximately 112. 
These examples illustrate how $\tau_1$ and $\tau_2$ affect the spectral compression process. 
Ideally, $\tau_1$ (related to the pump pulse duration) should match the roundtrip time $T_R$ as close as possible to concentrate energy primarily into diagonal peaks, while $\tau_2$ (related to the phase-matching bandwidth) determines the number of bins in the output JSI.

\begin{figure}[tb!]
\centering\includegraphics[trim=0 30 6 0,clip,width=3.5in]{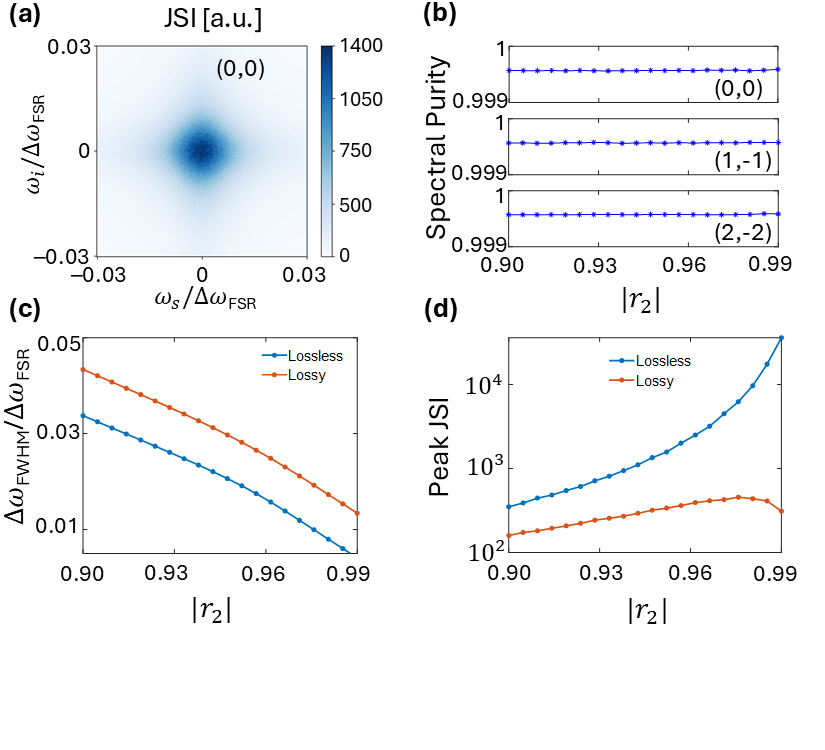}
\caption{Detaield biphoton characteristics for the case $(\tau_1,\tau_2) = (T_R,0.1T_R)$; $r_2 = 0.95$ is assumed. (a) Zoom-in of the peak at $(\omega_s,\omega_i)/\Delta\omega_\mathrm{FSR} = (0,0)$. (b) Spectral purity of the diagonal peaks located at $(\omega_s,\omega_i)/\Delta\omega_\mathrm{FSR} \in \left\{(0,0), (1,-1), (2,-2) \right\}$. (c) FWHM and (d) peak JSI of the central bin as functions of $|r_2|$, shown for both lossless (0~dB per roundtrip) and lossy (0.2~dB per roundtrip) cavities.}
\label{fig3}
\end{figure}

We now further investigate the characteristics of the output BFC, focusing on the first case in detail: $(\tau_1,\tau_2) =(T_R, 0.1T_R)$. 
Figure~\ref{fig3}(a) zooms in on the peak at $(\omega_s,\omega_i)/\Delta\omega_\mathrm{FSR} = (0,0)$. The peak exhibits a spectrally factorable form, with widths equal to the cavity linewidth along both the $\omega_s$ and $\omega_i$ axes. 
To quantify the factorability, we plot the spectral purity of the bins located at $(\omega_s,\omega_i)/\Delta\omega_\mathrm{FSR} \in \left\{(0,0), (1,-1), (2,-2) \right\}$ in Fig.~\ref{fig3}(b) as a function of output reflectivity $|r_2|$ for a cavity with no intrinsic loss. Spectral purity, the inverse of the number of effective factorable modes, is obtained via singular value decomposition of the wavefunction~\cite{christ2011probing,laudenbach2016modelling}. The spectral purity remains impressively high, approximately $0.9995$, across all considered $|r_2|$ values for all frequency-bin pairs. This level of purity, notably, also exceeds the maximum of $0.93$ typically achievable with BFC lines generated from fixed microring cavities~\cite{vernon2017truly, vaidya2020broadband, myilswamy2023time}.

Next, we present the full-width at half-maximum (FWHM) bandwidth $\Delta\omega_\mathrm{FWHM}$ and the peak JSI of the bin at $(\omega_s,\omega_i)/\Delta\omega_\mathrm{FSR} = (0,0)$ in Fig.~\ref{fig3}(c,d), respectively, for the lossless case (shown in blue) as functions of $|r_2|$. The FWHM is computed for the marginal signal spectrum and is the same for all frequency bins (results are identical for the idler). As $|r_2|$ increases, the biphoton spends more time inside the cavity, leading to greater spectral compression. 
In practice, intracavity losses need to be taken into account, which can be incorporated by replacing propagation terms of the form $-i\omega T_R$ in Eq.~(\ref{eq-H_W_W'}) by $-i\omega T_R -\alpha$ (similarly for $\omega^{\prime}$), where $2 \alpha$ represents the cavity roundtrip power loss. We then repeat the analysis for a cavity with a roundtrip loss of $0.2$~dB, a reasonable assumption for integrated cavities~\cite{myilswamy2020spectral}.  The spectral purity of the peaks remains unchanged for the considered values of $|r_2|$ (omitted for conciseness), while the FWHM and peak JSI of the central bin are shown in orange in Fig.~\ref{fig3}(c,d). 
The FWHM shows a similar trend to that of the lossless cavity but is higher due to the increased cavity linewidth. The peak JSI now exhibits a maximum of 455 at $|r_2|\approx0.98$, reflecting a tradeoff between spectral compression and loss; while spectral compression increases with $|r_2|$, the loss also increases, reducing the peak JSI, which nevertheless exceeds 100 for all considered values of $|r_2|$. This tradeoff mirrors the observations for single photons reported in our previous work~\cite{myilswamy2020spectral}. 

So far, we have assumed instantaneous switching of the input reflectivity from $0$ to $1$ at $t=T_R$. To explore the impact of finite rise time on spectral compression performance, we first note that 
an arbitrary time-varying input reflectivity $r_1(t)$ modifies the impulse response of Eq.~(\ref{eq-h_t_tau}) to
~\cite{crosignani1986time, sacher2008dynamics, myilswamy2020spectral}
\begin{equation}
\small
    \begin{aligned}
        & h(t,\tau) = \sum_{m=0}^{\infty} t_1(\tau)\left[ \prod_{n=0}^m b_n(\tau) \right] t_2 \delta \left(t-\tau - \left[m + \frac{1}{2}\right] T_R\right) \\
        \vspace{0.25in}
        & b_0(\tau) = 1; ~ b_n(\tau) = r_1(\tau + [m - n + 1]T_R)r_2  ~~\forall n \geq 1,
    \end{aligned}
    \label{eq-4}
\end{equation}
where $r_2^m$ in Eq.~(\ref{eq-h_t_tau}) has been replaced by factors $b_n(\tau)$ that describe the input mirror state for each bounce occurring at some integer multiple of $T_R$ after the field has entered the cavity.

\begin{figure}[tb!]
\centering\includegraphics[trim=0 0 0 0,clip,width=3.5in]{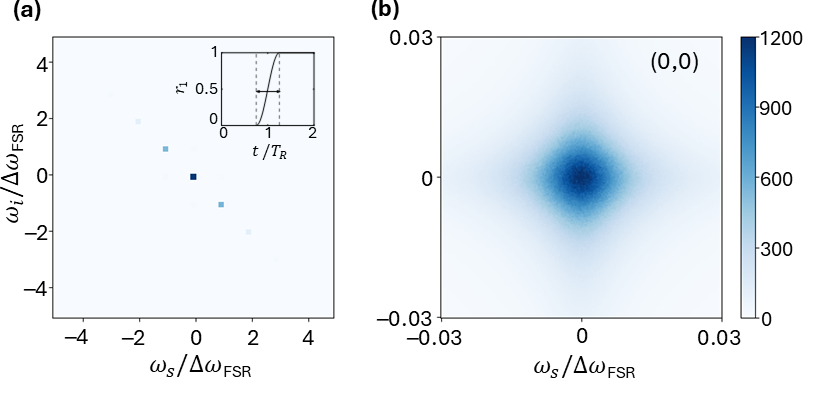}
\caption{Rise time considerations. The results are for $(\tau_1,\tau_2) = (T_R,0.1T_R)$, zero intrinsic cavity loss, and $r_2 = 0.95$. The reflectivity $r_1(t)$ is modeled as a raised cosine, as illustrated in the inset of (a), with the arrow indicating the rise time. (a)~Output JSI and (b)~peak at $(\omega_s,\omega_i)/\Delta\omega_\mathrm{FSR} = (0,0)$.}
\label{fig4}
\end{figure}

As a specific example, we model $r_1(t)$ as a raised cosine 
function, smoothly increasing from  $0$ to $1$ over the time-interval $\beta T_R$ with $\beta=0.75$, as illustrated in the inset of Fig.~\ref{fig4}(a). 
An exact analytical form of $H(\omega,\omega^{\prime})$ is not obtainable in this case but can be numerically calculated using the 2-D Fourier transform. 
We again consider $(\tau_1,\tau_2) = (T_R, 0.1 T_R)$ and perform the analysis for a lossless cavity for simplicity. In contrast to instantaneous switching, reflection losses are anticipated at the input as the reflectivity ramps up, which we partially compensate by shifting the biphoton center to $\tau_s=0.59T_R$ (found numerically to give maximum output flux). Figure~\ref{fig4}(a,b) displays the output JSI and a close-up of the peak at $(\omega_s,\omega_i)/\Delta\omega_\mathrm{FSR} = (0,0)$. The peak JSI of the central bin decreases to 1200, compared to 1400 for the rapid switching case shown in Fig.~\ref{fig2}(a), which is attributed to reflection losses at the input.
However, the spectral purity and FWHM remain nearly unchanged at $0.9994$ and $0.0210 \Delta\omega_{\mathrm{FSR}}$, respectively,  compared to $0.9995$ and $0.0207\Delta\omega_{\mathrm{FSR}}$ in the rapid switching scenario. This demonstrates that the spectral compression performance remains remarkably similar and still significantly outperforms that of a fixed cavity, even with the nonzero rise time. Additionally, with TFLN EOMs attaining bandwidths of around 100 GHz~\cite{boes2023lithium}, rise times around 5~ps should be possible (i.e., half of one period), which are fast enough to support $\beta=0.75$ for a 150~GHz FSR---%
implying the ability to generate BFCs with ultrahigh bin spacings with current technology.  

In summary, we have introduced a novel method for periodic spectral compression of broadband photon pairs into BFCs using linear time-varying cavities. These BFCs maintain cross-bin  entanglement while exhibiting high spectral purity within each bin---valuable for heralding pure single photons. Crucially, our approach does not trade off purity for brightness,  enhancing its value in the quantum networking toolkit. 
Furthermore, our theoretical analysis of the time-varying cavity using impulse response functions provides a versatile framework that can be applied to other contexts as well. With an increasing body of work exploring time-varying cavities to overcome fundamental limitations in electromagnetics~\cite{hayran2023using}, our spectral compression scheme is implementation-agnostic and applicable to any resonant system with switchable coupling, making it a valuable addition to the field.

\smallskip
\textbf{Acknowledgments.} J.A.G. thanks the John Martinson Honors College Undergraduate Research Fellowship at Purdue University. Parts of this work were presented at the Conference on Lasers and Electro-Optics (CLEO) 2024 as paper number FTu4F.1.

\textbf{Funding.} Air Force Office of Scientific Research (FA9550-20-1-0283); U.S. Department of Energy (ERKJ353).

\textbf{Disclosures.} The authors declare no conflicts of interest.

\textbf{Data availability.} Data underlying the results can be obtained from the authors upon reasonable request.

\end{document}